\documentclass[preprint,review,12pt]{elsarticle}
\usepackage{graphicx}
\usepackage{amssymb}
\usepackage{amsbsy}
\usepackage[version=3]{mhchem}
\usepackage[nodots]{numcompress}

\usepackage{times}
\usepackage{color}
\usepackage{graphicx}
\usepackage{amsmath,amsbsy,amsfonts,mathrsfs}
\newcommand{\be}{\begin{equation}}
\newcommand{\ee}{\end{equation}}
\newcommand{\bea}{\begin{eqnarray}}
\newcommand{\eea}{\end{eqnarray}}
\newcommand{\bg}{\begin{figure}}
\newcommand{\eg}{\end{figure}}
\newcommand{\bi}{\begin{itemize}}
\newcommand{\ei}{\end{itemize}}
\journal{Carbon}

\begin{document}

\begin{frontmatter}


\title{Effect of van der Waals forces on the stacking of coronenes encapsulated in a single-wall carbon nanotube and many-body excitation spectrum}

%\title{}


%%%%%%%%%%%%%%%%%%%%%%%%%%%%%%%%%%%%%%%%%%%%%%%%%%%%%%%%%%%%%%%%%%%%%%%%%%%%%%%%%%%%%%%%%%%%%%%%%%%%%%%%%%%%%%%%%%%%%%%%%%%%%%%%%%%%%%%

\author[1]{Yannick J. Dappe}

\author[2]{Jos\'e I. Mart\1nez}
\ead{joseignacio.martinez@uam.es}

\cortext[cor1]{Corresponding author}

\address[1]{Service de Physique de l'Etat Condens\'e, DSM/IRAMIS/SPEC, CEA Saclay URA CNRS 2464, F-91191 Gif-Sur-Yvette Cedex, France; 
CEA Saclay, DSM/IRAMIS/SPCSI, B\^atiment 462, F-91191 Gif sur Yvette, France}
\address[2]{Dept. de F\1sica Te\'orica de la Materia Condensada,
Universidad Aut\'onoma de Madrid, ES-28049 Madrid, Spain}

%%%%%%%%%%%%%%%%%%%%%%%%%%%%%%%%%%%%%%%%%%%%%%%%%%%%%%%%%%%%%%%%%%%%%%%%%%%%%%%%%%%%%%%%%%%%%%%%%%%%%%%%%%%%%%%%%%%%%%%%%%%%%%%%%%%%%%%

%%%%%%%%%%%%%%%%
\begin{abstract}
%%%%%%%%%%%%%%%%

We investigate the geometry, stability, electronic structure and optical properties of C$_{24}$H$_{12}$ coronenes encapsulated in a 
single-wall (19,0) carbon nanotube. By an adequate combination of advanced electronic-structure techniques, involving weak and van der 
Waals interaction, as well as many-body effects for establishing electronic properties and excitations, we have accurately characterized 
this hybrid carbon nanostructure, which arises as a promising candidate for opto-electronic nanodevices. In particular, we show that the 
structure of the stacked coronenes inside the nanotube is characterized by a rotation of every coronene with respect to its neighbors 
through van der Waals interaction, which is of paramount importance in these systems. We also suggest a tentative modification of the 
system in order this particular rotation to be observed experimentally. A comparison between the calculated many-body excitation spectrum 
of the systems involved reveals a pronounced optical {\it red-shift} with respect to the coronene-stacking gas-phase. The origin of this 
{\it red-shift} is explained in terms of the confinement of the coronene molecules inside the nanotube, showing an excellent agreement 
with the available experimental evidence.

%%%%%%%%%%%%%%
\end{abstract}
%%%%%%%%%%%%%%

% %%%%%%%%%%%%%%%%%%%%%%%%%%
% \begin{keyword}
% coronene stacking
% \sep
% carbon nanotubes
% \sep
% van der Waals
% \sep
% many body effects
% \sep
% excitation spectrum
% \end{keyword}
% %%%%%%%%%%%%%%%%%%%%%%%%%%


\end{frontmatter}

% \linenumbers



%%%%%%%%%%%%%%%%%%%%%%
\section{Introduction}
%%%%%%%%%%%%%%%%%%%%%%

In recent years, substantial effort has been devoted to the development of new nanostructures involving standard molecules, which states as a key concept for their applicability in
\textit{state-of-the-art} future nanodevices~[2-4].
%~\cite{Jaeger,Miyajima,Alves}. 
This burgeoning researching field gives rise to the building of more and more complex structures~\cite{Roquelet,Blau}, 
which are often inexpensive, versatile and stable materials that can traditionally be used at low temperatures, and are nowadays becoming attractive for high-temperature regimes. 
On this basis, the self-assembling of molecules through confinement within carbon nanotubes \cite{Kondratyuk,Sorin} offers a wide variety of interesting supramolecular structures. These systems are constituted 
of building blocks whose stability is ensured by weak intermolecular interactions. Among these mentioned interactions, the most fundamental and dominant one is the van der Waals attraction 
(vdW)~\cite{London1,London2}. It allows the assembling of molecules -- or groups of molecules -- together without substantially affecting the basic properties of each molecule individually. 
In fact, supramolecular chemistry~\cite{Yu}, as well as many biological systems~\cite{Kurita}, are based on these assembling of building blocks through vdW interaction. Nevertheless, 
despite the fundamental relevance of such interaction, its treatment is still very complicated to handle, especially from a first-principles framework. Some methods have been proposed 
in literature on the basis of Density Functional Theory (DFT)~[13-16],
%~\cite{Angyan05,Langreth05,Grimme,Lebegue10}, 
with relatively succesful results, but which are often limited by the size of 
the considered systems, and the computational resources.

In particular, the case of self-assembled stacked coronene molecules inside carbon nanotubes is of particular interest since: i) first, the coronene molecules can be considered as 
small pieces of graphene, and as such, present interesting electronic properties; ii) second, as planar molecules, coronene molecules can be easily arranged to form huge 1D-like 
supramolecular assembling. This property may be exploited for electronic transport purposes along the carbon nanotube axis; and iii) finally, as we shall revise in this work, the 
confinement of coronenes inside carbon nanotubes exhibits a non neglectable effect on their electronic structure. Consequently, these changes in the electronic structure may be utilized 
(and even tuned) by varying the size of the nanotube, in order to create new and efficient \textit{opto-electronic} devices.

In this communication, motivated by recent experimental results by Okazaki and coworkers~\cite{Okazaki}, we consider the case of $\pi$-stacked coronenes in a Single Walled Carbon Nanotube (SWCNT). 
With this study, we aim at giving new and deep insights in the structure and energetics, as well as the electronic properties, of such system. This complex has been partially characterized 
theoretically through standard DFT calculations in previous literature, but without considering the van der Waals interaction, which is of paramount importance for the intrinsic cohesion of 
this system. We shall show that a poor (or null) consideration of this challenging interaction leads to an erroneous description and characterization of this structure. On the other hand, 
from the electronic structure point of view, ``many-body'' techniques are dramatically necessary to properly describe the molecular gap, as well as the SWCNT band structure, -- and, 
therefore, the excitation spectrum --. To our knowledge, such an accurate framework has not been considered before.

In the following, in section 2 we will present our calculation methodology. We will recall our intermolecular perturbation theory plus DFT to determine the structural configuration of
the coronenes inside the nanotube. Then, we will present a ``many-body'' corrected DFT approach to determine accurate band-structure diagrams, as well as the excitation spectrum. In  
section 3 we will deeply discuss the structure obtained; and in section 4, we will provide the electronic structure results. All the previous results combined will be used to 
justify a very recent experimental evidence, by which an optical {\it red-shift} effect of the self-assembled C$_{24}$H$_{12}$ coronenes stacked inside the (19,0) SWCNT with respect to its 
diluted clean-phase is observed~\cite{Okazaki}. On the basis of our findings, future experimental work is proposed.

%%%%%%%%%%%%%%%%
\section{Method}
%%%%%%%%%%%%%%%%

%%%%%%%%%%%%%%%%%%%%%%%%%%%%%%%%%%%%%%%%%%%%%%%%%%%%%%%%%%%%%%%%%%%%%%%%%%%%%%%%%%%%%%%%%%%%%%%%%%%
\begin{figure}
\centerline{\includegraphics[width=\columnwidth]{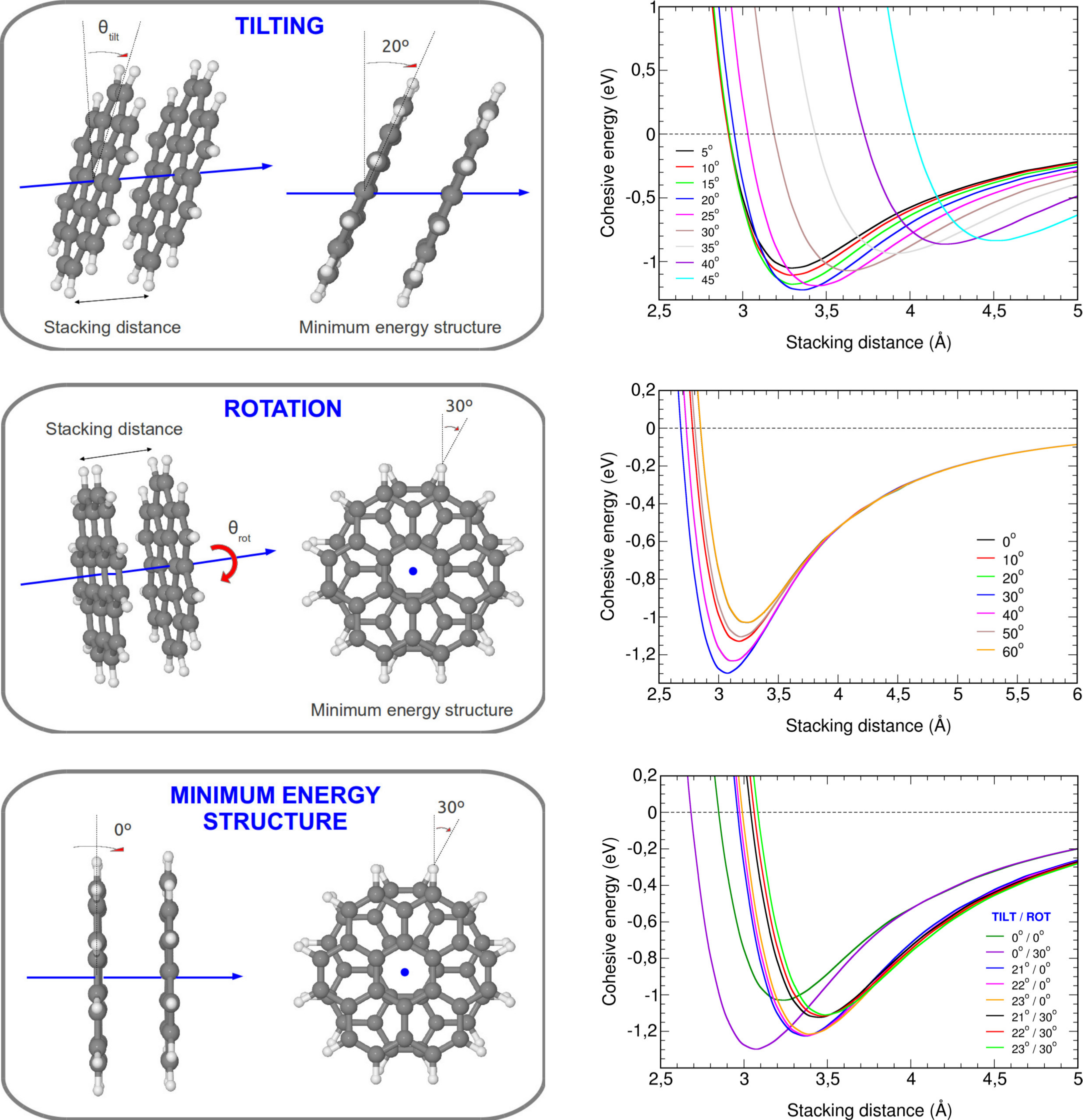}}
\smallskip \caption{(Color online) \textit{(Top panel right)} Cohesive energy of an isolated 
C$_{24}$H$_{12}$ coronene dimer as a function of the intermolecular stacking distance for several 
tilt angles calculated in the LCAO-$S^2$+vdW formalism. \textit{(Middle panel right)} Ibid for 
several roration angles with respect the axis dimer -- note that rotation angles of $40$, $50$ and 
$60$ degrees are respectively equivalent to rotation angles of $20$, $10$ and $0$ degrees --. 
\textit{(Bottom panel right)} Ibid for several tilt angles, considering a rotation angle of $0$ or 
$30$ degrees with respect to the dimer axis. Sketchs about tilt and rotation angles, and minimum 
energy structures are shown in left subpanels.} \label{P1}
\end{figure}
%%%%%%%%%%%%%%%%%%%%%%%%%%%%%%%%%%%%%%%%%%%%%%%%%%%%%%%%%%%%%%%%%%%%%%%%%%%%%%%%%%%%%%%%%%%%%%%%%%%

In order to shed some light on the interpretation of the mentioned optical {\it red-shift} evidence~\cite{Okazaki}, the geometrical and electronic structure, as well as the optical photoexcitation 
properties, of all these systems have been deeply analyzed from an accurate theoretical point of view by the combination of advanced theoretical techniques: the van der Waals interaction
for the geometry and energetics, and the ``many-body" effects for the effective correction of the electronic structure. On this basis, we have performed a full {\it state-of-the-art} theoretical 
characterization of all the relevant systems involved in the present study.

%%%%%%%%%%%%%%%%%%%%%%%%%%%%%%%%%%%%%%%%%
\subsection{The LCAO-$S^2$+vdW formalism}
%%%%%%%%%%%%%%%%%%%%%%%%%%%%%%%%%%%%%%%%%
 
As starting point, we first study the isolated coronene dimer, in order to properly describe, subsequently, the confinement of stacked coronenes inside the nanotube. All these systems 
are dominated by weak and van der Waals interactions, which will be accounted along the whole characterization process. For this purpose we have used the LCAO-$S^2$ + vdW~\cite{Dappe06,Dappe09}. 
This approach is based on the Density Functional Theory (DFT) in combination with an intermolecular perturbation theory to describe weak and van der Waals (vdW) interactions. Each 
interacting subsystem, namely in this case the coronene and the carbon nanotube, are calculated separately by DFT. The DFT computational scheme, as well as the theoretical foundations underlying 
our calculations -- a very efficient DFT localized orbital molecular dynamics technique ({\sc Fireball}) --, have been described in full detail elsewhere~[19-22],
%~\cite{Lewis01,Jelinek05-1,Jelinek05-2,Lewis11}, 
and here we will only summarize the main points. 

We first analyze all the systems involved in this study by using a self-consistent version of the Harris-Foulkes LDA functional~\cite{Harris,Foulkes} -- instead of the traditional 
Kohn-Sham (KS) functional based on the electronic density --, where the KS potential is calculated by approximating the total charge by a superposition of spherical charges around each atom. 
The {\sc Fireball} simulation package uses a localized optimized minimal basis set~\cite{Basanta06}, and the self-consistency is achieved over the occupation numbers through the Harris 
functional~\cite{Demkov95}. Besides, the LDA exchange-correlation energy is calculated using the efficient multicenter weighted exchange-correlation density approximation 
(McWEDA)~\cite{Jelinek05-1,Jelinek05-2}. To these DFT calculations we add ``weak interactions'', which can be seen as two opposite contributions. The first one, named ``weak chemical'' interaction 
is due to the small overlaps between electronic densities of the interacting subsystems. Therefore, this energy can be determined as an expansion of the wavefunctions and operators with 
respect to these overlaps. This expansion is based on a development in $S^2$ (since in the weak-interacting case the overlaps are really small) of the S$^{-1/2}$ term appearing in the L\"owdin 
orthogonalization, which induces a shift of the occupied eigenenergies of each independent subsystem, and leads to a repulsion energy between them. This shift can be defined as follows: 
\begin{eqnarray}
\nonumber \delta^S \varepsilon_n=
-\sum_m \frac{1}{2}
 \left[S_{nm}T_{mn}+ 
 T_{nm}S_{mn} \right] + 
\frac{1}{4}\sum_m 
| S_{nm}|^2(\varepsilon_n-\varepsilon_m ), \\
\delta^S \varepsilon_m=-\sum_n \frac{1}{2}
 \left[ S_{mn}T_{nm} + 
T_{mn}S_{nm} \right] + 
\frac{1}{4}\sum_n 
| S_{nm}|^2(\varepsilon_m-\varepsilon_n )
\label{overlap},
\end{eqnarray} where $S_{nm}$ and $T_{mn}$ are the overlapping and hopping integrals between eigenvectors $m$ and $n$ in the two interacting subsytems, respectively; {\it i.e.} each coronene 
molecule, or the SWCNT and the molecule. The effect of the hopping matrix elements, $T_{mn}$, can be calculated in a standard intermolecular second-order perturbation theory as:
\begin{eqnarray}
\nonumber\delta^T \varepsilon_n=
\sum_m \frac{|T_{mn}|^2}
{\varepsilon_n-\varepsilon_m}, \\
\delta^T \varepsilon_m=
\sum_n \frac{|T_{mn}|^2}
{\varepsilon_m-\varepsilon_n}.
\label{hopping}
\end{eqnarray}
Thus, we obtain the following ``weak chemical'' contribution to the interaction energy:
\begin{eqnarray}
E_{\textrm{weak chemical}}=2\sum_{n=occ.}(\delta^S\varepsilon_n+\delta^T\varepsilon_n) + 2\sum_{m=occ.}(\delta^S\varepsilon_m+\delta^T\varepsilon_m),
\label{one-electron}
\end{eqnarray} where a multiplicative factor of 2 has been included to take into account the spin degeneracy and only filled states are considered. Typically, in graphitic or hydrocarbonated 
materials this contribution is in general repulsive. 

The second contribution is the pure van der Waals interaction, which finds its origin in charge fluctuations in each subsystems, arising from oscillating dipoles whose interaction gives 
the attractive part of the cohesive energy. This interaction is treated in the dipolar approximation, and added in perturbations to the total energy of the system. The balance of the 
two contributions gives the equilibrium configuration of the system. This formalism~\cite{Dappe06,Dappe09} has already provided excellent results in the study of a wide range of graphitic 
materials~\cite{Savini11}, or encapsulated molecules in nanotubes~\cite{Debbichi12}.

%%%%%%%%%%%%%%%%%%%%%%%%%%%%%%%%%%%%
\subsection{Band-structure diagrams}
%%%%%%%%%%%%%%%%%%%%%%%%%%%%%%%%%%%%

Based on the previously obtained optimized vdW geometries, density functional theory (DFT) and ``many-body'' calculations -- via the quasi-particle Green function screened interaction 
approximation (GW)~\cite{Hybertsen}, combined in the next section to the Bethe-Salpeter equation (BSE)~[30-33]
%~\cite{Albrecht1,Albrecht2,Albrecht3,Onida} 
for the calculation of the excitation spectrum -- have been carried 
out on all the systems, in order to provide deep insights on their electronic structure (band-structure diagrams) and optical photoabsorption behavior (excitation spectrum). By accounting these
``many-body'' corrections, this study goes much deeper in the theoretical characterization of these systems than in any previous literature~\cite{Okazaki}.

DFT implementations used in practice have some well-known limitations due to the approximate description of exchange and correlation effects between the electrons, such as the known 
underestimation of the magnitude of the energy gap in semiconductors~\cite{Mowbray}. A similar error could be expected for the gap between occupied and unoccupied states (HOMO-LUMO)
gap in finite clusters and nanoparticles~\cite{Martinez}. Possible errors in the gap and in the asymptotic behavior of the XC potential in a semiconductor bulk could dramatically 
affect the energies of the excited electronic states and, thus, the band structure, photoabsorption spectrum or electron-phonon coupling. For this reason, the DFT results are effectively
corrected here with accurate electronic structure results obtained from ``many-body'' perturbation theory. To explore the influence of quasi-particle and electron-hole interaction effects 
in the band-structure diagrams of the different systems, we have used the GW (Green function - screened interaction approach)~\cite{Hybertsen}. For this purpose: i) as a starting point, 
we took the previously obtained vdW geometries and lattice parameters for establishing the ground-state electronic structure of all systems with the plane-wave {\sc PwScf} code~\cite{Baroni}, 
using for this step a Local Density (LDA) XC-functional~\cite{Perdew} -- it is well-known that most-improving quasi-particle corrections are obtained from local density electronic ground-states 
(see~\cite{Martinez} and references therein) --. Additionally, semicore-corrected norm-conserving pseudopotentials~\cite{Troullier-Martin1,Troullier-Martin2} were used to model the ion-electron interaction. 
The Kohn-Sham one-electron states were expanded in a series of plane-waves with an energy cutoff of 500 eV, and the Brillouin zone of all systems was $k$-sampled with ($12\times1\times1$) 
Monkhorst-Pack grids, guaranteeing a full convergence in energy and density. The self-consistent density was determined by iterative diagonalization of the Kohn-Sham Hamiltonian at k$_B$T = 
0.1 eV, using Pulay mixing of densities, and all total energies were extrapolated to k$_B$T = 0 eV. Spin-polarized calculation tests were also carried out to check the influence of the spin 
in these systems. Ground-state calculations for the vdW-based geometries converged fast, and revealed small forces on every atom ranging between $0.1$ and $0.2$ eV/\AA, which manifests an excellent
transferability of geometries between simulation codes. This justifies their use in the more accurate electronic structure calculations. Additionally, spin-polarized calculations
did not show either any significant variation with respect to the spin-unpolarized cases. Step ii) was performed with the {\sc Yambo} code~\cite{Marini}, and more than 200 unoccupied bands 
for each system were necessary to obtain self-consistent GW converged results. 

%%%%%%%%%%%%%%%%%%%%%%%%%%%%%%%%
\subsection{Excitation spectrum}
%%%%%%%%%%%%%%%%%%%%%%%%%%%%%%%%

A method extensively used to calculate the photoabsorption spectrum of nanostructures, surfaces and solids is the time-dependent density functional theory~\cite{Marques,Martinez06} (TDDFT), 
which states a good performance for low electronically correlated systems through the time-evolution of the Kohn-Sham states. Nevertheless, as previously indicated, the implementations used 
in practice have some well-known limitations due to the approximate description of exchange and correlation effects between the electrons. For this reason, here we go beyond standard TDDFT 
by applying many-body perturbation theory via the quasi-particle approximation GW correction~\cite{Hybertsen} to the standard DFT, combined to the Bethe-Salpeter 
equation~[30-33]
%~\cite{Albrecht1,Albrecht2,Albrecht3,Onida} 
for the calculation of the excitation spectrum, as follows: i) as a starting point, we consider the LGA+GW electronic quasi-particle corrections~\cite{Hybertsen} to the electronic structure 
for the ground-state of all the systems as calculated in the steps i) and ii) explained in the previous subsection; and ii) finally, the BSE~[30-33]
%~\cite{Albrecht1,Albrecht2,Albrecht3,Onida} 
was solved for coupled electron-hole 
excitations~\cite{Onida}, thereby accounting for the screened electron-hole attraction and the unscreened electron-hole exchange. Step ii) was also performed with the YAMBO code \cite{Marini}. 
This formalism, GW+BSE, includes, by construction, excitonic effects, which are more important in highly correlated systems, combining the excellent performance of the GW+LDA on the ground-state 
electronic structure calculations, with the high accuracy of the BSE for the calculation of excitations.

%%%%%%%%%%%%%$$$$%%%%%%%%%%%%%%%%%%%%%%
\section{vdW Geometries and Energetics}
%%%%%%%%%%%%%%%%%$$$$%%%%%%%%%%%%%%%%%%

The results of our study can be divided in two main aspects: i) first, we will consider the structural aspects of the C$_{24}$H$_{12}$ coronenes encapsulated in carbon nanotubes due 
to the weak and van der Waals interactions. Coronene molecules stack into coaxial columns inside the nanotube through commonly named $\pi$-stacking interaction that we interpret 
here as weak and van der Waals interactions. Consequently, they are confined in the nanotube leading to different electronic properties. As a preliminary step, we firstly present 
some results on the isolated coronene-coronene dimer interaction. In a second step, we will present the interaction of a stacked coronene chain encapsulated inside the (19,0) carbon 
nanotube. The chosen size of the nanotube corresponds to that experimentally observed by Okazaki and coworkers~\cite{Okazaki}, adequate to accomodate in a suitable way the coronene 
column. ii) In a second part, we will present the electronic properties determined by ``many-body''-corrected DFT and the GW + Bethe-Salpeter calculations. This framework will be used 
to determine the photoexcitation spectrum of an infinite clean coronene-stacking column, as well as the coronene column inside the nanotube. As a complement, previously we also present 
a detailed study of the band-structure of this new hybrid system in order to interpret the obtained photoexcitation results.

%%%%%%%%%%%%%%%%%%%%%%%%%%%
\subsection{Coronene dimer}
%%%%%%%%%%%%%%%%%%%%%%%%%%%

%%%%%%%%%%%%%%%%%%%%%%%%%%%%%%%%%%%%%%%%%%%%%%%%%%%%%%%%%%%%%%%%%%%%%%%%%%%%%%%%%%%%%%%%%%%%%%%%%%%
\begin{figure}
\centerline{\includegraphics[width=9cm]{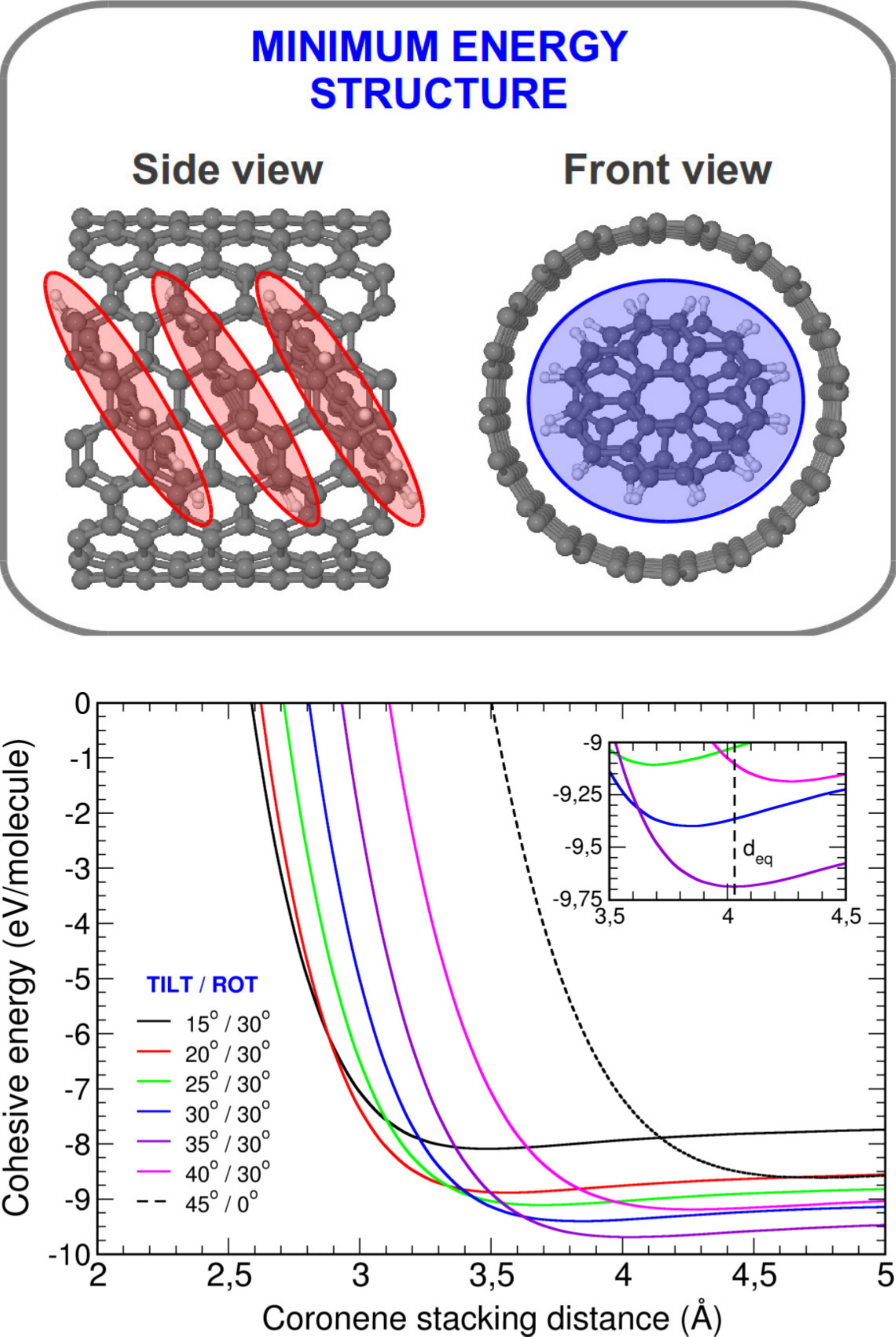}}
\smallskip \caption{(Color online). \textit{(Bottom panel)} Cohesive energy per coronene molecule 
(inserted between two neighboring coronenes) as a function of the intermolecular stacking distance
of an infinite chain of stacked C$_{24}$H$_{12}$ coronenes encapsulated inside the ($19,0$) SWCNT. 
This cohesive energy is depicted for several tilt angles, considering a rotation angle of $30$ 
degrees between the coronenes. The most stable structure (sketched in top panel) is obtained for a 
$35$ degrees tilt angle, at an intermolecular stacking distance of around 4~\AA. In bottom panel, 
total energy minimum region is shown as an enlarged-scale inset for a better visualization.} 
\label{P2}
\end{figure}
%%%%%%%%%%%%%%%%%%%%%%%%%%%%%%%%%%%%%%%%%%%%%%%%%%%%%%%%%%%%%%%%%%%%%%%%%%%%%%%%%%%%%%%%%%%%%%%%%%%

We have first considered the interaction between two coronene molecules forming an isolated dimer (without the nanotube), in order to determine the ``natural'' configuration of the dimer. 
All the calculations here have been performed using the LCAO-$S^2$ + vdW formalism explained above. In Figure~\ref{P1} (top panel) we have depicted the cohesive energy of the coronene dimer for 
various tilt angles as a function of the stacking distance between the molecules. The tilt angle is defined as a rotation angle of both molecules in parallel with respect to the $\hat{y}$-axis, 
as it can be seen in the top sketch of Figure~\ref{P1}. This specific angle will be considered with caution to define the coronene stacking-structure in the nanotube. We can observe a minimum energy 
of about -1.2 eV for the whole dimer, at a distance of 3.3~\AA, and for a tilt angle of 20 degrees. This result is rather different than for the coronene tilt angle in the monoclinic crystal, 
of around 46 degrees for a 3.4~\AA~intermolecular distance~\cite{Robertson}, since obviously we only consider an isolated coronene dimer. In a second calculation, we have determined, within the same 
formalism, the coronene dimer cohesive energy, considering the rotation of one molecule with respect to the other and with respect to the dimer-axis ($\hat{z}$-axis), at tilt angle
equal to 0. In this case, the minimum energy is found for a rotation angle of $30$ degrees, and for a distance of $3.1$ \AA, as represented in the middle panel of Figure~\ref{P1}. The geometrical 
configuration of the most stable situation is also represented in the middle sketch of Figure~\ref{P1}. This result is confirmed by previous studies, where the specific 30 degrees configuration is 
denominated as ``twisted-sandwich'' structure~\cite{Podeszwa}. In a previous DFT study~\cite{Obolensky}, by using a general-gradient {\sc B3Lyp} corrected functional, with no van der
Waals correction, this structure is not found, however, to be the most stable one, showing the limitation of the standard DFT approach to describe these systems. In a recent work by our group,
this essential role of the van der Waals interaction for the accurate determination of the geometrical structure has already been demonstrated for the C$_{60}$ fullerene confined in a carbon 
nanotube~\cite{Dappe11}. From a physical point of view, this stable situation corresponds to an $AB$-stacking configuration, similarly to the graphene case, whilst in the previous calculation, 
at zero tilt angle, it corresponds to an $AA$-stacking, which is more repulsive. This behaviour is corroborated by the obtained cohesive energy, which in this case is of around -1.3 eV.

Finally, in order to prepare the calculation of the stacked coronenes inside the carbon nanotube, we have calculated the isolated coronene dimer cohesive energy for several tilt angles, 
comparing in each case the situation with rotation angles of 0 and 30 degrees. Results are shown in bottom panel of Figure~\ref{P1}, where we can clearly appreciate that the most stable 
configuration corresponds to a zero tilt angle with $30$ degrees of rotation angle, obtaining the same stacking distance and energy as in the previous calculation. This seems to point out 
that the stacking of the coronenes inside the carbon nanotube will correspond to a deformation of this stable situation, due to the confinement within the nanotube walls. Therefore, in the 
next subsection, we will focus on this particular ``twisted-sandwich'' configuration to determine the coronenes packing inside the carbon nanotube.

%%%%%%%%%%%%%%%%%%%%%%%%%%%%%%%%%%%%%%%%%%%%%%%%%%%%%%%%%%%%%%%%%%%%%%
\subsection{Encapsulated coronene-stacking inside the carbon nanotube}
%%%%%%%%%%%%%%%%%%%%%%%%%%%%%%%%%%%%%%%%%%%%%%%%%%%%%%%%%%%%%%%%%%%%%%

In this subsection we present calculations of the encapsulation of stacked coronene molecules inside a ($19,0$) SWCNT. For this purpose, we have considered 3 coronenes inside the nanotube, 
as represented in the top sketch of Figure~\ref{P2}, and we focus on the cohesive energy of the coronene lying in the middle. Figure~\ref{P2} shows the evolution of this cohesive energy as a
function of the inter-coronene stacking distance for various tilt angles, and considering a rotation of 30 degrees with respect to the tube axis from one coronene to the other. 
This rotation angle has been chosen according to the previous results obtained for the single dimer calculation. The most stable structure is obtained for a tilt angle of 35 degrees, 
at an inter-coronene distance of 4~\AA. The corresponding cohesive energy is around of $9.7$ eV per coronene, with an excellent agreement with previous calculations on C$_{60}$ encapsulated 
inside a carbon nanotube~\cite{Dappe09}. This result is also in a good agreement with the experimental result giving a tilt angle of $\sim$23 degrees, as well as with the corresponding DFT 
calculations giving a tilt angle of 35 degrees and an inter-coronene distance of 4.2~\AA~\cite{Okazaki}. Nevertheless, as indicated above, the coronene stacking structure is rather different 
from the stability required by the system with a $30$ degrees rotation of each coronene with respect to its neighboring molecules. In order to check the validity of this result, we have also 
calculated the same structure within our formalism, with no rotation angle between the coronenes. In this case we were not able to determine a stable configuration in the system, since the 
lowest energy is obtained for a 45 degrees tilt angle. We guess that in such a case, the most stable configuration should be obtained for a tilt angle approaching to 90 degrees, which is 
equivalent to have no stacking between coronene molecules, but still forming chains. Of course, this result does not correspond at all to the Physics observed experimentally, meaning that 
the stable configuration requires a rotation of each coronene with respect to the other. Therefore, our study reveals a major aspect of coronene stacking encapsulated in carbon nanotubes, 
which is the 30 degrees rotation angle of each coronene with respect to the others. This aspect is even more important, since it has not been observed in this experiment. In addition 
to the calculation with a 30 degrees rotation, we represent in Figure~\ref{P2} the cohesive energy of the same system with no rotation, and for a tilt angle of 45 degrees, which is the lowest 
energy we obtained in this case. It is noticeable the existing moderately high energy difference of around $1$ eV per molecule with respect to our total minimum, taking into account the 
30 degrees rotation angle, which validates our model.

The discrepancy between our equilibrium structure and the one obtained in~\cite{Okazaki}, especially regarding the 30 degrees rotation angle resulting in our calculations, may be explained 
in terms of two different factors: i) first, our treatment takes into account weak and van der Waals interaction, which was not the case in the previous study, based on standard DFT calculations, 
and ii) second, in~\cite{Okazaki}, the authors have maintained a constant distance of 4.2~\AA~ between the coronenes in the nanotube, not allowing further lattice optimization. 
In the calculations we performed on this system, with no rotation angle, at a fixed distance of 4~\AA, we also determine that the 35 tilt angle yields the minimum energy of the system, 
eventhough this is not the absolute minimum (see Figure~\ref{P2}). By varying the distance, one can observe that in this configuration, increasing the tilt angle will increase the cohesive energy, 
probably until the coronenes stand all aligned inside the carbon nanotube. Consequently, our study goes beyond in the accurate characterization of this system, since we found that coronenes 
can only be stabilized by rotating each coronene with respect to its neighbors by $30$ degrees. Since it has not been observed experimentally, due essentially to the low response of hydrogen 
atoms in electron microscopy, we suggest future experimental work based on the observation by high-angle annular dark field scanning transmission electron microscopy 
(HAADF-STEM)~\cite{krivanek,Suenaga} of a similar system with some fluorine atoms substituting hydrogen atoms in the coronenes, as an example. One could consider stacked coronenes where some
hydrogen atoms could be subtituted by Fluorine atoms, in order to be observed by microscopy techniques, in such a way to be able of discriminating (and categorically determining) the exact rotation 
angle between coronenes in their ground-state configuration. We have studied such system within DFT and observed that it is perfectly stable in the nanotube, without altering any
of the structural properties. This procedure could be applied to a large amount of molecules stacked inside carbon nanotubes, since this rotation is probably 
a general behaviour in these systems.

%%%%%%%%%%%%%%%%%%%%%%%%%%%%%%%%%%%%%%%%%%%%%%%%%%%%%%%%%%%%%%%%%%%
\section{``Many-body'' Electronic Structure and Excitation Spectra}
%%%%%%%%%%%%%%%%%%%%%%%%%%%%%%%%%%%%%%%%%%%%%%%%%%%%%%%%%%%%%%%%%%%

%%%%%%%%%%%%%%%%%%%%%%%%%%%%%%%%%%%%%%%%%%%%%%%%%%%%%%%%%%%%%%%%%%%%%%%%%%%%%%%%%%%%%%%%%%%%%%%%%%%
\begin{figure}
\centerline{\includegraphics[width=\columnwidth]{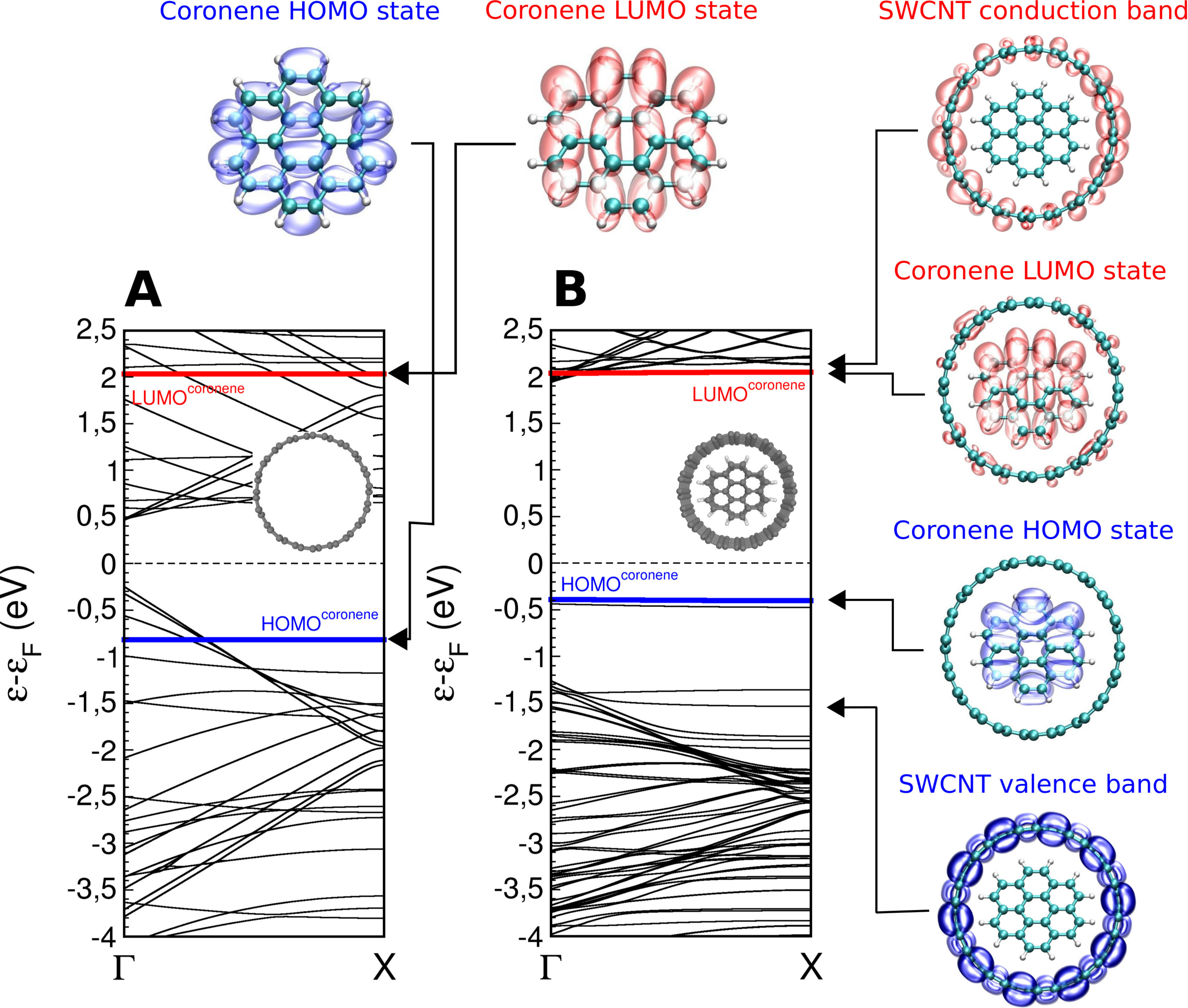}}
\smallskip \caption{(Color online) Quasi-particle GW-corrected DFT band-structure diagrams 
(referred to the Fermi energy level) for: A) Isolated ($19,0$) SWCNT. HOMO and LUMO levels of the 
self-assembled interacting C$_{24}$H$_{12}$ coronenes are also shown superimposed onto the nanotube 
band diagram (solid blue and red lines, respectively); and B) (19,0) SWCNT$@$C$_{24}$H$_{12}$ 
interacting system, where HOMO and LUMO coronene states have been visually emphasized within the 
band diagram (solid blue and red lines, respectively), along the most relevant symmetry line from 
$\Gamma \rightarrow$ X (along nanotube axis). Additionally, 3D orbital isodensities, all of them 
with a value of $2\times 10^{-5}$ e$^{-}$/\AA$^{3}$, are shown for the most representative states of 
the systems.} \label{P3}
\end{figure}
%%%%%%%%%%%%%%%%%%%%%%%%%%%%%%%%%%%%%%%%%%%%%%%%%%%%%%%%%%%%%%%%%%%%%%%%%%%%%%%%%%%%%%%%%%%%%%%%%%%

In this section we will present the results of electronic structure calculations, as well as excitation spectrum, of an infinite coronenes chain encapsulated in the carbon nanotube, as 
to be compared with the infinite coronenes chain in its clean dilute-phase. For the isolated infinite coronenes chain (dilute phase) we have taken the most stable structure obtained for 
the dimer, periodically extended to mimic the experimental dimensionality, and with the optimum stacking distance obtained in the previous section. Nevertheless, the geometry taken for 
the simulation of the infinite coronenes chain encapsulated inside the nanotube is slightly different from the most stable one obtained from the exhaustive vdW analysis. Due to the high 
computational requirements of our GW+BSE formalism, for this system we have taken the ground-state structure for the coronenes encapsulated inside the nanotube, with no accounting the 30 
degrees rotation angle. This rotation angle cannot be considered because the performance of our accurate formalism is just limited to one coronene per unit cell. The staking distance is 
that obtained in the geometrical analysis section. However, from our experience and regarding the next results, the rotation angle together with the electrostatic vdW interaction should 
not have a significant influence in the electronic structure scheme of the whole interacting system.

\subsection{Band-structure diagrams}

In Figure~\ref{P3}A we show the Quasi-particle GW-corrected DFT band-structure diagram (referred to the Fermi energy level) for the isolated infinite ($19,0$) zig-zag SWCNT for the most 
important symmetry line $\Gamma \rightarrow X$ (along nanotube axis). The LDA+GW band diagram reveals a value for the band gap of $\varepsilon_{g}^{(19,0)} = 0.76$ eV at the 
$\Gamma$ {\bf {\it k}}-point, to be directly compared with those shown in previous literature of $0.58$ and $0.62$ eV obtained by less accurate methods: {\it i.e.}, third-order nonlinear 
optical response \cite{Margulis} and standard DFT by using a GGA-B3LYP hybrid potential \cite{Matsuda}, respectively. To our knowledge, no experimental band-gap value is available for this 
nanotube. As known, the obtained band gap manifests the narrow band-gap semiconducting character of the ($19,0$) zig-zag SWCNT (for more details see~\cite{Margulis,Matsuda} and 
references therein). In Figure~\ref{P3}A, we also show superimposed on the band diagram the HOMO and LUMO levels of the self-assembled stacked C$_{24}$H$_{12}$ coronenes in their clean diluted phase 
(solid blue and red lines, respectively), keeping the same geometry and stacking periodicity they exhibit inside the nanotube fix. The obtained HOMO-LUMO gap has a value of
$\varepsilon_{g}^{(19,0)} = 2.92$ eV, where the HOMO and LUMO levels (referred to the Fermi energy level) lie inside the unoccupied and occupied bands of the isolated tube, 
respectively, before the interaction between both systems is produced. For completion, 3D orbital HOMO and LUMO isodensities (both with a value of $2\times 10^{-5} e^{-}$/\AA$^{3}$) are depicted in Figure~\ref{P3}. The 3D isodensity profiles for the self-assembled 
stacked C$_{24}$H$_{12}$ coronenes visibly show a higher localization of the orbital HOMO charge around the double bonds inside the coronenes,  whilst it is appreciable how the local symmetry 
of the LUMO completely changes with respect to the latest, making an optical transition between them permitted through the dispersion between the hybridized $\pi$-orbitals of neighboring 
coronenes, as explained in detail the spectroscopic analysis of~\cite{Martinez}. This analysis shall be extended in the excitation spectrum section.
 
Figure~\ref{P3}B shows the Quasi-particle GW-corrected DFT band-structure diagram (referred to the Fermi energy level) for the ($19,0$) SWCNT$@$C$_{24}$H$_{12}$ interacting system, where HOMO and LUMO 
coronene states have been visually emphasized within the band diagram (solid blue and red lines, respectively), along the most relevant symmetry line from  $\Gamma \rightarrow X$ (along 
nanotube axis). As an interesting result of the interaction between both subsystems, the LDA+GW band diagram reveals a global dilatation of all the bands around the Fermi energy, where a 
compression of occupied and empty bands is produced opening the original semiconducting band gap of the tube up to a value of $\varepsilon_{g}^{(19,0)} = 3.4$ eV at $\Gamma$ point. Now, 
the interacting HOMO and LUMO levels of the self-assembled stacked C$_{24}$H$_{12}$ coronenes (again solid blue and red lines) appear in the band diagram, but still, as expected, with an almost 
inexistent dispersion with the {\bf {\it k}} points, mainly due to the low vdW interaction between them, and also with the wall of the tube. Nevertheless, this apparently weak interaction 
of the internal coronenes with the tube is certainly responsible of the nanotube gap-opening and the compression of the states in the global band-structure diagram. It is noticeable how the 
coronene-phase LUMO state remains located just like before the interaction is produced with respect to the Fermi energy, but, after the interaction, it interestingly overlaps in energy with 
the new nanotube conduction band. However, no apparent hybridization is produced between both states, each of them keeping its original coronene / nanotube character (justified below). 
Meanwhile, the coronene-phase HOMO has moved up towards the Fermi energy level, by globally reducing their original HOMO-LUMO gap up to a value of $\varepsilon_{g}^{cor} = 2.52$ eV. Other 
interesting effect we can emphasize is that apparently the {\bf {\it k}}-dispersion profile of the nanotube bands seems not to show any variation before and after the interaction is produced, 
which manifest the robustness of the nanotube electronic structure, and the weak chemical interaction between the coronenes and the nanotube. Thus, summarizing, the interaction between the 
stacked coronenes inside the ($19,0$) nanotube produces a three-folded effect: i) an opening of the nanotube band gap, losing the semiconducting character; ii) a compression of the occupied 
and unoccupied bands around the Fermi level; and iii) a closing of the coronene-phase HOMO-LUMO gap, showing a noticeable hybridization between the coronene-phase LUMO state and the former 
nanotube conduction band. 

Similarly to the previous case, 3D isodensity profiles for the SWCNT$@$C$_{24}$H$_{12}$ system are also shown for the most relevant bands / states: former 
nanotube valence and conduction bands, and HOMO and LUMO states of the coronene-phase (all of them with a value of $2\times 10^{-5} e^{-}$/\AA$^{3}$). As appreciable in Figure~\ref{P3}, the 3D 
isodensities associated to the nanotube valence and conduction band after the interaction is produced reveal a location of the orbital charge completely surrounding the tube. Regarding the 
same value taken in all cases for the depiction of the 3D isosurfaces, this justifies our previous statement, when we concluded no hybridization at all between the valence or conduction bands 
with any of the coronene-phase states; in particular, between the new nanotube conduction band and the coronene LUMO state which are located (not hybridized) at the same energy with respect 
to the Fermi level. Once again, 3D coronene-phase HOMO and LUMO orbital isodensities show a higher localization of the orbital HOMO charge around the double bonds inside the coronenes, with 
no variations with respect to their isolated diluted case, which indicated that the interaction between the coronenes and the nanotube is weak. In the same way, orbital LUMO isodensity profile 
keeps unaltered as well its symmetry after the interaction between the nanotube and the coronenes is produced. Giving the similar symmetry of the HOMO and LUMO states than for the 
non-interacting case, this interpretation for the interacting ($19,0$) SWCNT$@$C$_{24}$H$_{12}$ system still makes an optical transition between them permitted. Nevertheless, the optical 
transition will be produced, in this case, at slightly lower excitation energy, since the HOMO-LUMO gap has been reduced by around $\sim 0.4$ eV. This analysis shall be extended in the 
excitation spectrum section. 	 

\subsection{Excitation spectra}

%%%%%%%%%%%%%%%%%%%%%%%%%%%%%%%%%%%%%%%%%%%%%%%%%%%%%%%%%%%%%%%%%%%%%%%%%%%%%%%%%%%%%%%%%%%%%%%%%%%
\begin{figure}
\centerline{\includegraphics[width=\columnwidth]{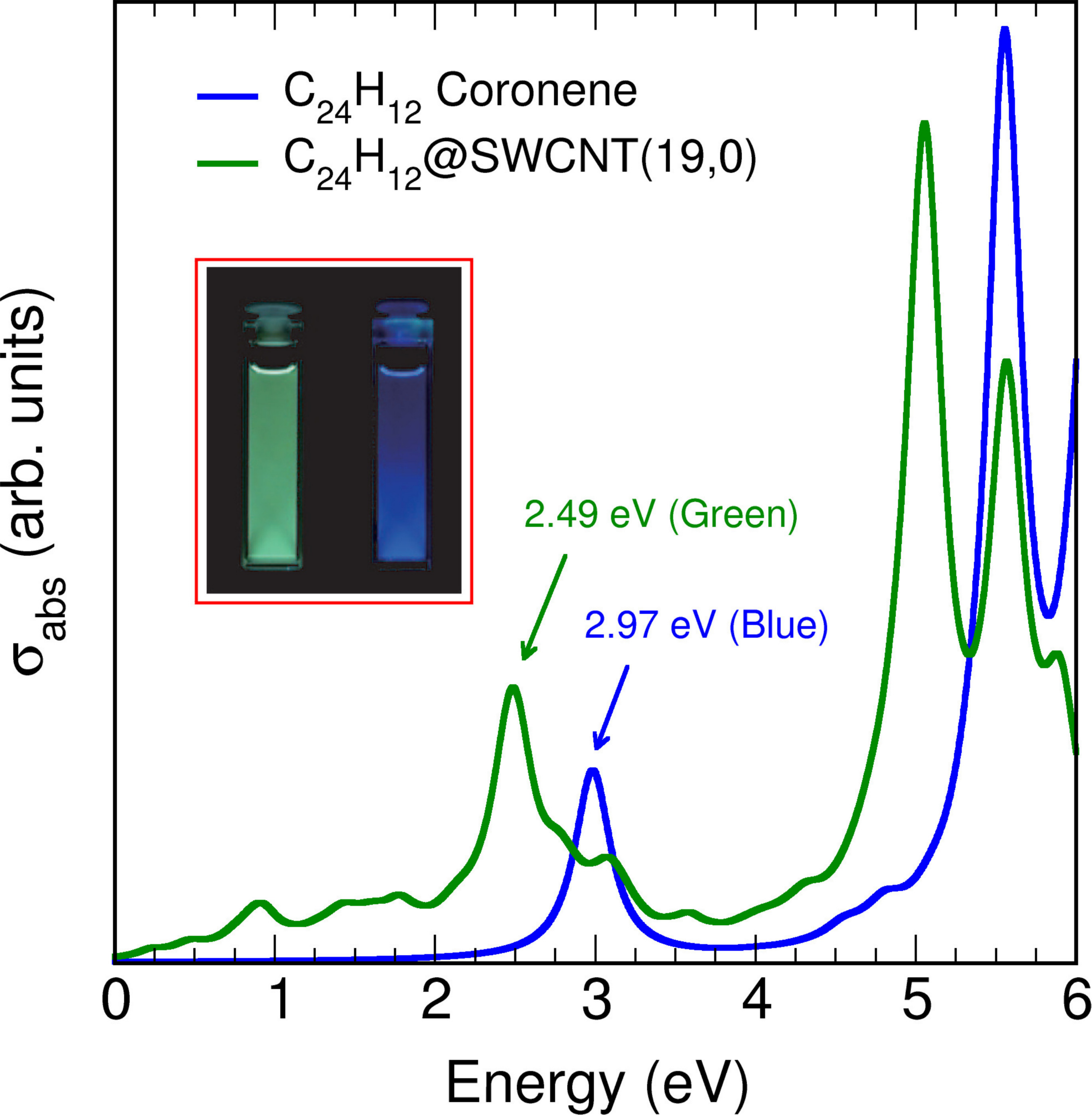}}
\smallskip \caption{(Color online) GW-BSE photoabsorption cross sections (in arbitrary units) as a 
function of the excitation energy (in eV) for: i) diluted self-assembled C$_{24}$H$_{12}$ coronenes 
phase (solid blue line); and ii) ($19,0$) SWCNT$@$C$_{24}$H$_{12}$ interacting system (solid green 
line). In order to mimic the experimental resolution~\cite{Margulis} the excitation peaks have been 
broadened by Lorentzian functions. The corresponding geometrical structure is shown in Figure~\ref{P2}. 
Photograph of the fluorescence from coronenes$@$SWCNTs in SDBS micelle solution (inset: left) and 
coronene in $n$-hexane solution (inset: right) under UV lamp irradiation (lex=$356$ nm)
~\cite{Okazaki}.} \label{P4}
\end{figure}
%%%%%%%%%%%%%%%%%%%%%%%%%%%%%%%%%%%%%%%%%%%%%%%%%%%%%%%%%%%%%%%%%%%%%%%%%%%%%%%%%%%%%%%%%%%%%%%%%%%

Figure~\ref{P4} shows the GW-BSE photoabsorption cross sections (calculated as explained above) as a function of the excitation energy for both the diluted self-assembled C$_24$H$_12$ coronenes 
phase (solid blue line), and the ($19,0$) SWCNT$@$C$_{24}$H$_{12}$ interacting system (solid green line) up to $6$ eV. The finite width of the excitation peaks in an experimental spectrum 
- linked to the accessible resolution - is mostly determined by the temperature. In our calculations, on the other hand, each excitation peak in the spectrum has been broadened by a 
Lorentzian profile, where the broadening parameter $\eta$ is set equal to $0.2$ eV, a value commonly used in order to mimic the experimental resolution~\cite{Martinez06}.

Regarding the figure, no significant photoabsorption signal is observed up to around $2$ eV for both systems; nevertheless, the low-energy absorption regime (up to $2$ eV) is slightly 
more pronounced for the SWCNT$@$C$_{24}$H$_{12}$ interacting system, which may be due to the (weak) interaction between the stacked coronenes and the nanotube, and the dispersion of extended
states coming from the periodic tube. The spectra of the self-assembled C$_{24}$H$_{12}$ coronenes phase and the ($19,0$) SWCNT$@$C$_{24}$H$_{12}$ interacting system up to $4$ eV are characterized 
by a main peak centered at $2.97$ and $2.49$ eV, respectively. According to the band diagrams shown in Figure~\ref{P3}, this peak seems to correspond to a permitted optical transition (see 
the orbital-based analysis below) from the HOMO state to the LUMO state of the staked coronenes chain (isolated or confined inside the nanotube). The intensity of this pronounced peak is 
slightly higher for the case of the coronenes inside the nanotube, which means that the electronic transition probability from HOMO to LUMO states increases for the confined coronenes. 
Nevertheless, both spectra profiles are quite similar, except for a relative {\it red-shift} of about $0.48$ eV of the confined coronenes with respect to their dilute phase. Besides, 
both also show a very similar photoabsorption profile for the high absorption energy range (from $4.5$ eV), arising from less bounded states weaker in energy.

The experimentally observed {\it red-shift} between the excitation spectra, roughly introduced in the previous section, is responsible of the change of the optical light-color emission
from blue to green (see inset in Figure~\ref{P4}). Photons of selected energies responsible to excite electrons from the HOMO to the LUMO states in the stacked coronenes, decay subsequently by 
the emission of blue or green light. As explained, the interaction between the stacked coronenes inside the ($19,0$) nanotube produces a compression of the occupied and unoccupied bands 
around the Fermi level, and a closing of the coronene-phase HOMO-LUMO gap from $2.97$ eV (corresponding to a blue-light emission) to $2.49$ eV (corresponding to a green-light emission). 
The origin of this optical {\it red-shift} from one system to the other can be justified in base to a detailed orbital-based analysis, which allows us to properly identify and label the 
HOMO and LUMO coronene-phase states (see isosurfaces in Figure~\ref{P3}), respectively, as $e_{2g}$ and $e_{2u}$ in D$_{6h}$ symmetry (in spectroscopic nomenclature), and thus the transition 
between these states is permitted, but only allowed through {\bf {\it z}}-axis polarized light (with $a_{2u}$ symmetry) - in this case along the longitudinal stacking axis -, while it 
is forbidden through {\bf {\it x}} and {\bf {\it y}}-axis polarized light (with $e_{1u}$ symmetry). This interesting fact is also confirmed directly by the GW-BSE photoabsorption results, 
in which the {\bf {\it z}}-axis polarization gives the only contribution to the lowest energy peaks in spectra of Figure~\ref{P4}. Moreover, it can be easily checked that there are no other 
allowed transition between the HOMO of the coronene-phase and the frontier orbitals, apart from the mentioned permitted $e_{2g} \rightarrow e_{2u}$ transition (see more details about 
this analysis in~\cite{Martinez} and references therein). This transition is produced in both systems at different HOMO-LUMO gap energies of the coronene-phase, depending on if they 
are in the clean dilute phase, or confined inside the nanotube, which is manifested in the blue and green colored-light emissions observed in the experiment. \\
Obviously, all these results as well as the van der Waals structures determined previously are sensitive to the particular choice of nanotube that we study here. In a future work we can
think about a systematic study of coronene encapsulation in nanotube with respect to the chirality and the size of the host nanotube.  

%%%%%%%%%%%%%%%%%%%%%
\section{Conclusions}
%%%%%%%%%%%%%%%%%%%%%

Following the experiment of Okazaki {\it et al.}~\cite{Okazaki}, we have studied theoretically the encapsulation of coronenes in a carbon nanotube, focusing on the 
($19,0$) SWCNT$@$C$_{24}$H$_{12}$ system. In a first part, using a DFT-based formalims which takes into account weak and van der Waals interactions, we have studied the coronene dimer as 
a previous step for an accurate description of stacked coronenes encapsulated in the carbon nanotube. The major result that we obtain is a rotation of $30$ degrees of each coronene with 
respect to its neighbors along the nanotube axis, leading to a tilt angle in the nanotube of $35$ degrees and an intermolecular distance of $4$~\AA. Actually this rotation between molecules,
which has not been seen experimentally, is abolutely necessary to obtain the stability of the system. Whereas in previous DFT calculations (without any van der Waals interaction), at fixed
intermolecular stacking distance, the general configuration is very similar to our result, the $30$ degrees rotation to ensure stability has not been shown before. In a second step, 
starting from this determined configuration, we have used the GW formalism to accurately calculate the bandstructure of the whole ($19,0$) SWCNT$@$C$_{24}$H$_{12}$ system and compared it 
with the isolated ($19,0$) SWCNT. As a main result, we can clearly observe a compression of the coronene levels as well as an opening of the SWCNT gap, due to the mutual interaction 
between the two systems. Moreover, a precise determination of the excitation spectrum shows a clear {\it red-shift} from $2.97$ eV in the coronene gas phase to $2.49$ eV for the
encapsulated system. This result is in excellent agreement with the experimental results by Okazaki {\it et al.}~\cite{Okazaki}, confirming the transition of the photoemission spectrum 
from blue to green emitted light. It can be deduced also that even if the stacking of coronene is a requisite for the stability of the structure ({\it i.e.} through the $30$ degrees rotation 
angle and the consequent tilt angle), the drastic change in electronic structure is purely due to the interaction with the nanotubes and the band compression. Finally, we suggest here an 
experiment to observe this fundamental rotation between coronenes in the nanotube (which is probably a general trend in such molecular systems), by substituing in each coronene some of the 
hydrogen atoms by Fluorine atoms for example, these last ones being more observable in transmission electron microscopy.

\bigskip

\noindent {\bf Acknowledgements}

\bigskip

This work is supported by Spanish MICIIN (grant FIS2010-16046), the CAM (grant S2009/MAT-1467), and the European Project MINOTOR (grant FP7-NMP-228424). JIM acknowledges funding 
from MICIIN through JdC Program.


%%%%%%%%%%%%%%%%%%%%%%%%%%%

%%%%%%%%%%%%%%%%%%%%%

%%%%%%%%%%%%%%
\end{document}